\documentclass[twocolumn]{aastex62}
\usepackage{amsmath}

\received{, 2021}
\revised{, 2021}
\accepted{, 2021}

\submitjournal{ApJ}

\shorttitle{Properties of Long and Short Fast Radio Bursts}
\shortauthors{Li et al.}

\begin{document}

\title{Long and Short Fast Radio Bursts are Different from Repeating and Non-repeating Transients}
\author[0000-0001-6469-8725]{X. J. Li}
\affiliation{School of Physics and Physical Engineering, Qufu Normal University, Qufu 273165, China}
\author[0000-0001-6469-8725]{X. F. Dong}
\affiliation{School of Physics and Physical Engineering, Qufu Normal University, Qufu 273165, China}
\author[0000-0001-6469-8725]{Z. B. Zhang$^{\dag}$}
\affiliation{School of Physics and Physical Engineering, Qufu Normal University, Qufu 273165, China}
%\email{astrophy0817@163.com}
\email{astrophy0817@163.com;dili@nao.cas.cn}
\author[0000-0003-3010-7661]{D. Li$^{\ddag}$}
%\email{dili@nao.cas.cn}
\affiliation{National Astronomical Observatories, Chinese Academy of Sciences, Beijing 100101, People¡¯s Republic of China}
\affiliation{NAOC-UKZN Computational Astrophysics Centre, University of KwaZulu-Natal, Durban 4000, South Africa}
%%%%%%%%%%%%%%%%%%%%%%%%%%%%%%%%%%%%%%%%%%%%%%%%%%%%%%%%%%%%%%%%%%%%%%%%%%%%%%%%
%%%%%%%%%%%%%%%%%%%%%%%%%%%%%%%%%%%%%%%%%%%%%%%%%%%%%%%%%%%%%%%%%%%%%%%%%%%%%%%%
\begin{abstract}
We collect 133 Fast Radio Bursts (FRBs), including 110 non-repeating and 23 repeating ones, and systematically investigate their observational properties.
To check the frequency dependence of FRB classifications, we define our samples with a central frequency below/above 1GHz as subsample I/II. We find that there is a clear
bimodal distribution of pulse width for the subsample I. And If we classify FRBs into short FRBs (\emph{s}FRBs) ($<$100 ms) and long FRBs (\emph{l}FRBs) ($>$100 ms) as done for short and long Gamma-Ray Bursts (GRBs), the \emph{s}FRBs at higher central frequency are commonly shorter than those at lower central frequency not only for non-repeating but also repeating \emph{s}FRBs.
Secondly, we find that fluence and  peak flux density are correlated with a power law relation of $F \varpropto S{^{\gamma}_{p,obs}}$ for both \emph{s}FRBs and \emph{l}FRBs whose distributions are obviously different. Thirdly, the \emph{l}FRBs with isotropic energies ranging from $10^{42}$ to $10^{44}$ erg are more energetic than the \emph{s}FRBs in the $F- DM_{EX}$ plane, indicating that they are two representative types. Finally, it is interestingly note that the peak flux density behaves an independence on the redshift when the distance of the FRBs becomes far enough, which is similar to the scenario of peak flux evolving with redshift in the field of GRBs. We predict that fainter FRBs at higher redshift of $z>2$ can be successfully detected by FAST and SKA in the near future.

\end{abstract}

\keywords{
\href{http://astrothesaurus.org/uat/739}{High energy astrophysics (739)};
%\href{http://astrothesaurus.org/uat/1338}{Radio astronomy (1338)};
\href{http://astrothesaurus.org/uat/2008}{Radio transient sources (2008)};
\href{http://astrothesaurus.org/uat/1339}{Radio bursts (1339)};
\href{http://astrothesaurus.org/uat/508}{Extragalactic radio sources (508)};
\href{http://astrothesaurus.org/uat/1340}{Radio continuum emission (1340)}
%\href{http://astrothesaurus.org/uat/1766}{Very Large Array (1766)};
}

%%%%%%%%%%%%%%%%%%%%%%%%%%%%%%%%%%%%%%%%%%%%%%%%%%%%%%%%%%%%%%%%%%%%%%%%%%%%%%%%
%%%%%%%%%%%%%%%%%%%%%%%%%%%%%%%%%%%%%%%%%%%%%%%%%%%%%%%%%%%%%%%%%%%%%%%%%%%%%%%%
\section{Introduction} \label{sec:intro}
Fast Radio Bursts (FRBs) discovered by the Parkes 64-m Telescope for the first time are transient radio pulses of millisecond
durations that flash randomly in the sky \citep{Lorimer2007,Thornton2013}. Their isotropic peak
luminosities are widely distributed from about 10$^{38}$ erg s$^{-1}$ to
10$^{46}$ erg s$^{-1}$ and the typical isotropic energies vary from about 10$^{35}$ erg to 10$^{43}$ erg
\citep{Zhang2018a,Zhang2020}.
%As the number of FRBs increases, the study of observational properties, propagation effect, radiation mechanism, population classification, etc. becomes important in the FRB field.

FRB 121102 as the first repeater was reported by \cite{Spitler2016} and 10 repeating bursts discovered by Arecibo Observatory subsequently. Then, \cite{Scholz2016} detected six follow-up bursts from this source: five bursts with the Green Bank Telescope (GBT) at 2 GHz, and one at 1.4 GHz with the Arecibo Observatory. Till now, FRB 121102 has been found to reburst a few thousand times \citep{Li2019,Li2021}. In the year of 2018, Canadian Hydrogen Intensity Mapping Experiment \citep[CHIME,][]{CHIME/FRB2018} reported the second source of repeating FRB 180814.J0422+73 \citep{CHIME/FRB2019a}. Currently, there are already about two dozens of repeaters published in literatures (see Table 1 of this work). As the number of repetition increases, some statistical studies of the observational properties of these repeating bursts have been conducted. For instance, it is found that the
waiting time distribution of FRB 121102 shows a clear bimodal distribution and does not correlate with
the burst intensity, suggesting some external mechanisms for these repeating bursts \citep{Li2019,Sun2021}.
Furthermore, \cite{Kumar2019} detected two repetitions of FRB 171019 with the Green Bank Telescope (GBT), the brightness of which
is a factor of $\sim$ 590 fainter than the first detection of FRB 171019 in the Australian Square Kilometre Array Pathfinder (ASKAP) fly's eye survey \citep{Shannon2018}. Meanwhile, \cite{Luo2020} carried out four follow-up observations of FRB 180301 firstly discovered by the Parkes radio telescope \citep{Price2019}, using the Five-hundred-meter Aperture Spherical radio Telescope (FAST) \citep{Li2018}, and found 15 repetitions from FRB 180301. Clearly, whether all FRBs repeat is an open question. Especially, a few previous ``non-repeating'' FRBs have been identified to be repeating events. Therefore, restudying the statistical features of FRBs with a complete sample becomes more and more urgent and necessary.
%it is not clear whether all non-repeating FRBs repeat because of the absence of the follow-up observation or the insufficiently telescope sensitivities.

At present, various progenitor models are proposed to explain FRBs, most of which involve compact objects \citep{Platts2019} \footnote{https://frbtheorycat.org}. Some source models are catastrophic and can only be applied to explain non-reapeating FRBs, such as the mergers or interactions of compact binaries \citep{Kashiyama2013,Totani2013,Mingarelli2015,Bhattacharyya2017,Dong2018}, collapse of compact objects \citep{Falcke2014,Zhang2014}, collisions of asteroids with neutron stars \citep{Geng2015,Huang2016}. After repeating FRB 121102 were detected, many models for repeating FRBs have been developed. A leading model for repeating FRBs is extragalactic magnetars \citep{Lyubarsky2014,Beloborodov2017,Metzger2019}. In addition, \cite{Gu2016} reported that the interaction between the bipolar magnetic fields of a neutron star and a magnetic white dwarf can be considered as a possible origin of repeating FRBs. \cite{Dai2016} proposed that a strongly magnetized NS encountering an extragalactic asteroid belt (EAB) around a stellar-mass object can arise a repeating FRB \citep{Dai2020}. Besides, there are also more other models, for example, neutron star cosmic combs \citep{Zhang2017,Zhang2018b} and young rapidly rotating pulsars can also lead to repeating FRBs \citep{Lyutikov2016}. Very recently, \cite{Geng2021} proposed that the repeating FRBs can be produced by the intermittent fractional collapses of the crust of a strange star and the 16-day periodicity of FRB 180916.J0158+65 can be well interpreted.

The observed spectral-temporal differences support the different origin between repeating and non-repeating FRBs.
Generally, the pulses of non-repeating FRBs are shorter in duration than those of repeating FRBs \citep{Scholz2016,CHIME/FRB2019b,Fonseca2020}. Compared to non-repeating FRBs, repeating FRBs show complex subpulse frequency structure and drifting and spectral
variation \citep{CHIME/FRB2019a}. Polarimetric measurements exhibit diversity, including a constant polarization angle for some repeating FRBs or variable polarization angles for some non-repeating FRBs \citep{Masui2015,Michilli2018,CHIME/FRB2019b,Cho2020,Luo2020}.
How the FRBs populate is always one of the most exciting open questions.
Considering the above controversial results, we expand the FRB samples to include the latest FRBs published in recent papers in order to disclose the nature of these radio transients in statistics reliably and systematically.
%For comparison, we divide them into two groups named as short and long FRBs
Using a larger repeating FRB sample, we deeply explore the possible  observational differences between repeating and non-repeating FRBs. It is worth pointing out that we define the short and long FRBs, and compare their observational parameters for the first time in this work. In addition, we examine the detectability of FRBs by different radio telescopes at higher redshift. The sample selection and data preparation are given in Section 2. The results are presented in Section 3. Finally, we summarize and discuss the results in Section 4.
%%%%%%%%%%%%%%%%%%%%%%%%%%%%%%%%%%%%%%%%%%%%%%%%%%%%%%%%%%%%%%%%%%%%%%%%%%%%%%%%
%%%%%%%%%%%%%%%%%%%%%%%%%%%%%%%%%%%%%%%%%%%%%%%%%%%%%%%%%%%%%%%%%%%%%%%%%%%%%%%%
\section{DATA and Method} \label{sec:observations}

Our sample comprises 133 FRBs, of which 129 FRBs are taken from the database of FRB Catalogue \citep[published up to July 2020, refer to][]{Petroff2016}, a nearby repeating FRB 20200120E on the outskirts of M81 was detected by the CHIME \citep{Bhardwaj2021}, and three new non-repeating FRBs, 181017.J0036+11, 181118 and 181130 were reported by FAST \citep{Niu2021}.
In total, there are 110 non-repeating FRBs and 23 repeating FRBs (including 116 repeaters) in our samples. Table 1 lists the first pulses of 23 repeating FRBs with the observed width ($W_{obs}$), redshift (z), DM, peak flux density ($S_{p,obs}$), fluence (F) and the references therein.

Generally, the total DM of a FRB at a redshift z is contributed by four components, i.e.
 \begin{equation} \label{equ1}
  \begin{split}
    DM & =DM_{MW}+DM_{EX}  \\
    & = {DM_{MW}+DM_{IGM}+\frac{DM_{host}+DM_{src}}{1+z}},
 \end{split}
 \end{equation}
where $DM_{EX}$ is an excess of DM with respect to the Milky Way value of $DM_{MW}$, the subscripts MW,
IGM, host, src refer to the contributions of the plasma from the Milky Way, intergalactic medium, FRB host galaxy and source environment, respectively \citep{Zhang2018a,Xiao2021} .

The isotropic peak luminosity and isotropic energy of a FRB at a central frequency of $\nu_c$ from \cite{Zhang2018a} can be calculated as
 \begin{equation} \label{equ2}
  \begin{split}
    L_p & \simeq4\pi D_L^2 S_{\nu,p} \nu_c  \\
    & =(10^{42} erg s^{-1}) 4\pi (\frac{D_L}{10^{28}cm})^2 \frac{S_{\nu,p}}{Jy} \frac{\nu_c}{GHz},
 \end{split}
 \end{equation}
 \begin{equation} \label{equ3}
  \begin{split}
    E & \simeq\frac{4\pi D_L^2}{1+z}F_\nu \nu_c  \\
    & =(10^{39} erg) \frac{4\pi}{1+z} (\frac{D_L}{10^{28}cm})^2 \frac{F_\nu}{Jy\cdot ms} \frac{\nu_c}{GHz},
 \end{split}
 \end{equation}
where $S_{\nu,p}$ is the specific peak flux density (in unit of Jy), $F_\nu$ is the
fluence (in unit of Jy ms), ${D_L}$ is the luminosity distance (in unit of cm), and $\nu_c$ is the central frequency (in unit of GHz). It is worth noting that \cite{Aggarwal2021} argued that the bandwidth other than the central frequency should be utilized to estimate the isotropic energies of repeating FRBs especially \citep [see also][]{Petroff2016}. We nonetheless use central frequency as the parameter since the majority of FRB samples detected by multiple telescopes within different bandwidths are non-repeating ones currently.
Throughout this paper, a flat $\Lambda$CDM universe with $\Omega_m$=0.286, $\Omega_\Lambda$ = 0.714 and H$_0$ = 69.6 km s$^{-1}$ Mpc$^{-1}$ has been assumed \citep{Bennett2014}.

%=====================================================================
\begin{deluxetable*}{llllccccc}
%\tablenum{}
\tablecaption{Parameters of the first-detected events for the 23 repeating FRBs \label{tab1}}
\tablewidth{0pt}
\tabletypesize{\footnotesize}
\tablehead{
\colhead{No.}&\colhead{FRB}& \colhead{DM}&\colhead{SNR} & \colhead{z}&\colhead{W$_{obs}$}
& \colhead{S$_{p,obs}$} & \colhead{F}&\colhead{Ref}\\
\colhead{}&\colhead{}&\colhead{(pc cm$^{-3}$)}&\colhead{} & \colhead{} & \colhead{(ms)} & \colhead{(Jy)} & \colhead{(Jy ms)}&\colhead{}
}
\startdata
1&20200120E	      &87.782$\pm$0.003&22.9   &$<$0.03  & 0.16$\pm$0.05  &1.8$\pm$0.9    &    2.25$\pm$0.12&[1]\\
2&190907.J08+46	  &310.0$\pm$0.4&10$^e$       &0.21       & 3$\pm$1       &1.7$\pm$0.6    &	0.3$\pm$0.2    &[2]\\
3&190604.J1435+53 &552.6$\pm$0.2&33.8$^e$	      &0.43       & 3$\pm$0.4    &0.9$\pm$0.4    &    8.3$\pm$2.8 &[2]\\
4&190417.J1939+59 &1378.1$\pm$0.2&13.4$^e$   &1.08       & 3.3$\pm$0.9    &4.4$\pm$0.8    &	0.5$\pm$0.2&[2]\\
5&190303.J1353+48 &221.8$\pm$0.5&11.2$^e$     &0.16       & 2.0$\pm$0.3    &0.5$\pm$0.3    &	2.3$\pm$0.9&[2]\\
6&190222.J2052+69 &460.6$\pm$0.1&31.2$^e$      &0.31       & 2.97$\pm$0.9   &1.9$\pm$0.6$^a$&    7.5$\pm$2.3&[3]\\
7&190212.J18+81   &301.7$\pm$0.3&9.9$^e$      &0.21       & 4.1$\pm$1.6    &0.4$\pm$0.3    &	3.0$\pm$1.5&[2]\\
8&190213.J02+20   &651.1$\pm$0.4&9.9$^e$      &0.51       & 10$\pm$2	   &0.06$\pm$0.03$^d$ &	0.6$\pm$0.3 &[2]\\
9&190209.J0937+77 &424.6$\pm$0.6&11.8$^e$      &0.32       & 3.7$\pm$0.5    &0.4$\pm$0.2$^a$&	2.0$\pm$1.0&[3]\\
10&190208.J1855+46&579.9$\pm$0.2&12.2$^e$      &0.42       & 0.91$\pm$0.16  &1.4$\pm$0.6    &	0.4$\pm$0.2  &[2]\\
11&190117.J2207+17&393.3$\pm$0.1&24.2$^e$      &0.29       & 1.44$\pm$0.03 &1.7$\pm$0.6    &	5.9$\pm$1.6  &[2]\\
12&190116.J1249+27&444.0$\pm$0.6&12.9$^e$     &0.35       & 4.0$\pm$0.5    &0.3$\pm$0.2$^a$&	0.8$\pm$0.4&[3]\\
13&181128.J0456+63&450.2$\pm$0.3&23.4$^e$      &0.28       & 2.43$\pm$0.16  &0.5$\pm$0.3$^a$&	4.4$\pm$2.2&[3]\\
14&181119.J12+65  &364.2$\pm$1.0&11.1$^e$      &0.28       & 6.3$\pm$0.6    &0.3$\pm$0.2$^a$&	1.8$\pm$0.8&[3]\\
15&181030.J1054+73&103.5$\pm$0.7&11.5$^e$      &0.05       & 0.59$\pm$0.08  &3.2$\pm$1.7$^a$&	7.3$\pm$3.8&[3]\\
16&181017.J1705+68&1281.9$\pm$0.4&12.9$^e$      &1.03       & 13.4$\pm$1.4   &0.4$\pm$0.3$^a$&   1.0$\pm$0.5&[3]\\
17&180916.J0158+65&349.2$\pm$0.4&18.7$^e$                   &0.12       & 1.40$\pm$0.07  &1.4$\pm$0.06$^a$&   2.3$\pm$1.2&[3]\\
18&180908.J1232+74&195.7$\pm$0.9&10.4$^e$                 &0.13       & 1.91$\pm$0.1   &0.6$\pm$0.4    &   2.7$\pm$1.1&[2]\\
19&180814.J0422+73&189.38$\pm$0.1&24    &0.09       & 2.6$\pm$0.2    &8.08$\pm$5.80$^d$ &   21$\pm$15 &[4]\\
20&180301	      &522$\pm$5&20	                  &0.37      & 2.18$\pm$0.06  &1.2$\pm$0.1    &	1.3&[5]\\
21&171019	      &460.8$\pm$1.1&23.4                  &0.35       & 5.4$\pm$0.3    &40.5$^c$       &	219$\pm$5   &[6]\\
22&121102	      &557.4$\pm$2.0&14                  &0.31       & 3.0$\pm$0.5    &0.4${^{+0.4}_{-0.2}}$& 1.2${^{+1.6}_{-0.55}}$&[7]\\
\hline
23&151125$^b$         &273$\pm$4&8.5      &0.19       & 1680$\pm$40$^c$&0.54          &    2450 &[8]\\
\enddata
\tablecomments{ \\
Ref-
[1] \cite{Bhardwaj2021}; [2] \cite{Fonseca2020}; [3] \cite{CHIME/FRB2019b};
[4] \cite{CHIME/FRB2019a}; [5] \cite{Price2019}; [6] \cite{Shannon2018}; [7] \cite{Spitler2016}; [8] \cite{Fedorova2019b}.\\
$^a$ The parameters are not given in FRB Catalog (https://frbcat.org) but given in literatures.\\
$^b$ A very long repeating \emph{l}FRB.\\
$^c$ The parameter is not given in literatures but given in FRB Catalog.\\
$^d$ These values are estimated with $F= S_{p,obs}\times W_{obs}$.\\
$^e$ These parameters are not reported in both literatures and FRB Catalog but reported in CHIME/FRB Catalog 1 (https://www.chime-frb.ca/catalog).
}
\end{deluxetable*}
%=====================================================================

\section{Results} \label{sec:results}

\subsection{Pulse width} \label{sec:agn}
There are 103 non-repeating and 111 pulses for 23 repeating FRBs with width measured. A Kolmogorov-Smirnov (K-S) test to repeaters and non-repeaters returns the statistic D = 0.22 and P = 0.01 for the $W_{obs}$ distributions. If adopting a critical value D$_\alpha$=0.19 at a significance level of $\alpha$= 0.05, one can conclude that the width distributions of repeating FRBs are somewhat different from non-repeating ones.

%When 11 \emph{l}FRBs are deducted, the average $W_{obs}$ of the 109 repetitions of the repeating FRBs (short repeating FRBs) and 93 non-repeating FRBs (short non-repeating FRBs) are 5.25${^{+0.18}_{-0.15}}$ ms and 3.94${^{+0.09}_{-0.09}}$ ms, respectively. The results show that the non-repeating FRBs are shorter in duration than the repeating FRBs, which is consistent with the previous conclusions \citep{Scholz2016,CHIME/FRB2019b,Fonseca2020,Bhattacharya2020}.
\begin{figure*}
\begin{center}
\plottwo{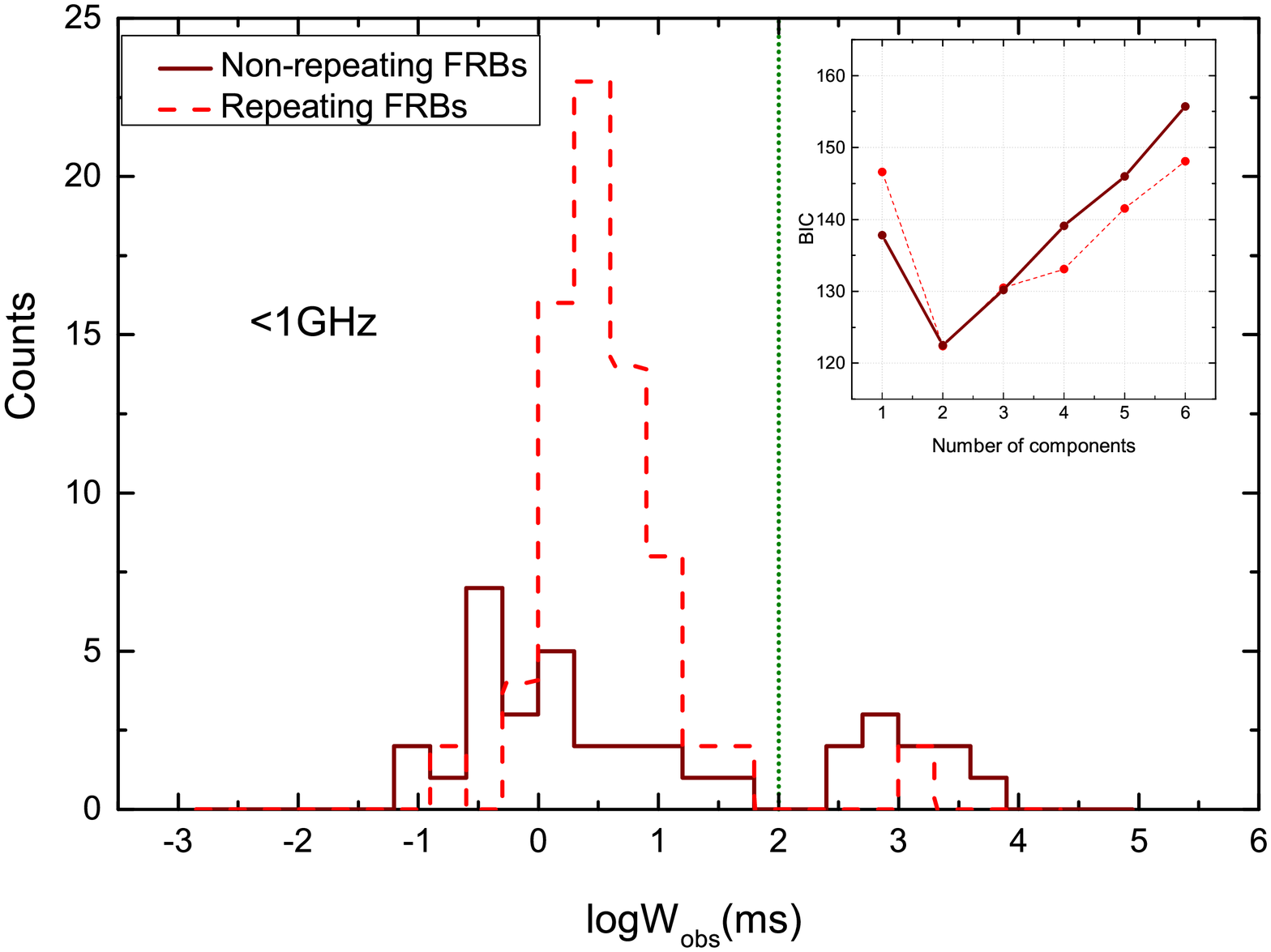}{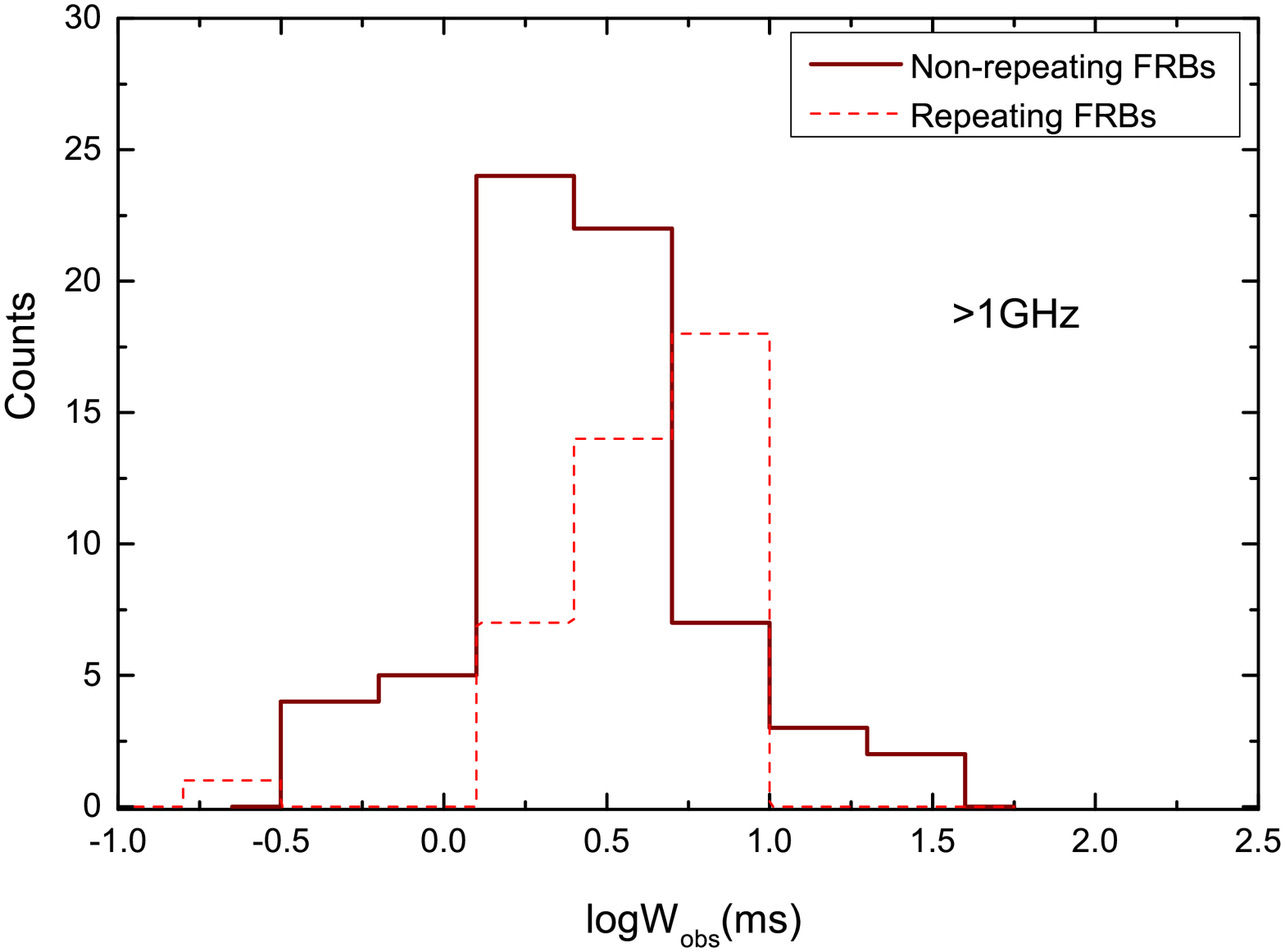}
\end{center}
\caption{Distributions of $W_{obs}$ for non-repeating bursts (solid) and repeating bursts (dashed) detected at a central frequency below 1GHz (left panel) and above 1GHz (right panel). Short and long FRBs are divided by the dotted green vertical line at $W_{obs}$ =100 ms phenomenologically. The insert in the left panel shows the inferred BIC values versus the number of components. }
\label{fig2}
\end{figure*}
%=====================================================================

%=====================================================================
To check the frequency dependence of FRB classifications, we define the
FRBs with a central frequency below 1 GHz as subsample I and the FRBs with a central frequency above
1 GHz to be subsample II, temporally.
%we divide the repeating and the non-repeating FRBs into two subsamples according to the center frequency, respectively.
In Figure \ref{fig2}, the left panel shows the $W_{obs}$ distributions of subsample I at lower frequency and the right panel shows for the subsample II at higher frequency. Then, we perform the maximum likelihood (ML) analysis along with Gaussian mixture model (GMM) to find the number of components in terms of Bayesian Information Criterion (BIC) for subsample I with PYTHON machine learning package SCIKIT-LEARN \citep{Pedregosa2011}. The BIC value is plotted against the component number in the inset of left panel like some previous works  \citep{Kass1995,Zhangzb2016,Tarnopolski2016}. It is interestingly found that the BIC curves of distinct kinds of FRBs always reach the smallest value at the point of two components. Additionally, the BIC differences between 1 and 2 components are $\triangle$BIC = 15 and 24 for non-repeating and repeating FRBs, respectively. This evidently demonstrates that the lognormal widths of both non-repeating and repeating FRBs in our lower (unlike higher) frequency sample are bimodally distributed. Motivated by the similarity to the duration distributions of GRBs, we separate FRBs into two classes, that is short FRBs (\emph{s}FRBs) with duration less than 100 ms and long FRBs (\emph{l}FRBs) with duration longer than 100 ms. It is found that there is no \emph{l}FRB events in the subsample II.
The average $W_{obs}$ of the repeating and non-repeating \emph{s}FRBs are 5.74${^{+0.17}_{-0.17}}$ ms and 4.53${^{+0.45}_{-0.37}}$ ms in subsample I (4.60${^{+0.09}_{-0.01}}$ ms and 3.96${^{+0.21}_{-0.15}}$ ms in subsample II), respectively. The average $W_{obs}$ of 10 non-repeating \emph{l}FRBs is 1698$\pm11.31$ ms. For the two bursts of repeating \emph{l}FRB 151125, the values of $W_{obs}$ are 1680$\pm40$ ms and 1470$\pm40$ ms, comparable to the average value of non-repeating \emph{l}FRBs.  As a whole, the duration of \emph{l}FRBs is about three orders of magnitude longer than that of \emph{s}FRBs for both repeating and non-repeating FRBs. Note that all \emph{l}FRBs except 151125 are ``one-off'' events in our sample.
%It is interesting that the same result goes for the \emph{l}FRBs.

In addition, we find that the pulses in non-repeating \emph{s}FRBs are narrower than those in repeating ones not only at higher frequency but also at lower frequency as shown in Figure \ref{fig2}. However, the result of \emph{l}FRBs is ambiguous, which is likely biased by a small number of \emph{l}FRBs especially for non-repeating \emph{l}FRBs. Interestingly, the ultra-long radio pulses with a duration of 1-2 seconds have been merely detected at lower frequency of $<$ 1GHz. Furthermore, we find that non-repeating \emph{s}FRBs with higher central frequency is shorter than those of non-repeating ones with lower central frequency, and so do repeating \emph{s}FRBs, similar to the conclusion reported by \cite{Gajjar2018} and \cite{Josephy2019}, indicating the scatter broadening phenomenon when the signal
propagates through the interstellar medium \citep{Xu2016,Nikitin2021}.
%the degree of variability is not obvious.
%In general, the results indicate that the two classes of \emph{s}FRBs might originate from different physical processes in view of the observed width with different frequency observations at least.

\subsection{The densities of flux and fluence} \label{sec:agn}
\begin{figure}
\begin{center}
\includegraphics[width=1\linewidth]{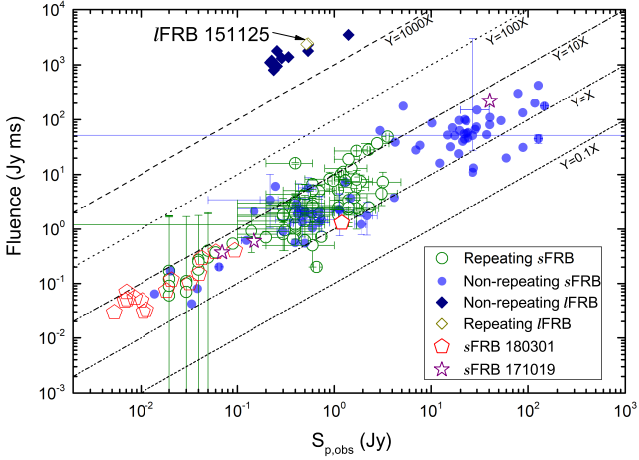}
\end{center}
\caption{Fluence is plotted against S$_{p,obs}$. Five ratios of fluence to $S_{p,obs}$ are symbolized by the dashed, dotted, dash-dotted, dash-dotted-dotted, and short dashed lines for 1000, 100, 10, 1 and 0.1, respectively.
\label{fig3}}
\end{figure}

Figure \ref{fig3} shows that non-repeating \emph{s}FRBs typically have higher brightness than repeating bursts on average. However, the distributions of their intensities are overlapped each other. Meanwhile, it is found that the range of non-repeating \emph{s}FRBs is wider than that of repeating bursts, and the intensities of repeating FRBs are relatively lower. In Figure \ref{fig3}, it is noticeable that two repeating \emph{s}FRBs 171019 (purple stars) and 180301 (red pentagons) had been wrongly dealt to be one-off events. Follow-up observations indicate that their radiative intensities became fainter and fainter since the first pulse. Considering the two special FRBs without fluence reported in FRB Catalogue and literatures, we have chosen the relation $F= S_{p,obs}\times W_{obs}$ \citep{Petroff2016} to estimate the unknown fluences.

Unlike before, we propose a power law correlation of $F=\mu S{_{p,obs}^\gamma}$ to apply for all kinds of FRBs. Table 2 lists our best fitting results and correlation coefficients. The results demonstrate that the power law relations exist in all kinds of FRBs with a roughly consistent power-law index of $\gamma\sim0.9$. However, the intercept of the $F-S_{p,obs}$ relations of \emph{l}FRBs and \emph{s}FRBs are obviously different. Figure \ref{fig3} also shows that most \emph{s}FRBs are located in the region of $F \sim (1 - 10) S_{p,obs} $. It is surprisingly found that the \emph{l}FRBs lie in the area of $F > 10^3 S_{p,obs}$ and they violate the $F-S_{p,obs}$ relation of \emph{s}FRBs in evidence.
Very interestingly, the repeating \emph{l}FRB 151125 is consistent with those non-repeating \emph{l}FRBs, which may hint that some ``one-off'' \emph{l}FRBs currently could be repeative subsequently.

\subsection{Fluence versus dispersion measure} \label{sec:agn}
\begin{figure}
\begin{center}
\includegraphics[width=1\linewidth]{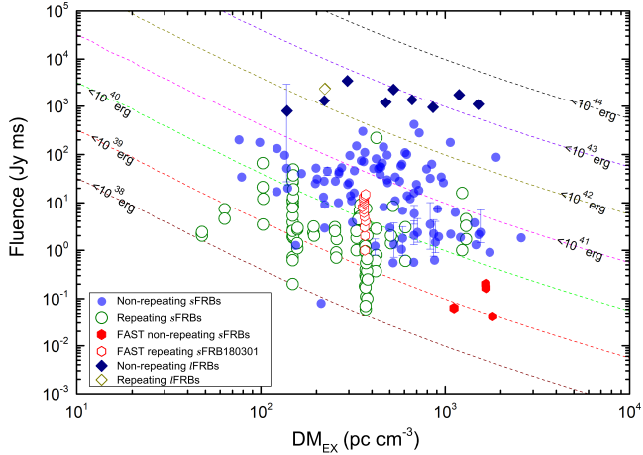}
\end{center}
\caption{The Fluence-$DM_{EX}$ diagram. The dashed curves labeled with individual numbers stand for the up-limits of radio isotropic energies of FRBs from Equation (\ref{equ3}).
\label{fig4}}
\end{figure}

Figure \ref{fig4} shows that the updated fluence-$DM_{EX}$ relation after the \emph{l}FRBs and some new bursts are added. Note that $DM_{EX}$ equals to $DM_{IGM}$ under the assumption of $DM_{host}=0$ and $DM_{src}=0$.
A rough relation $z\sim DM_{IGM}/855$ $pc$ $cm^{-3}$ had been used to estimate the redshift of a FRB \citep{Zhang2018a}. A higher fluence range of the \emph{l}FRB samples can be distinguished clearly. Given that \emph{s}FRB samples, our result is basically consistent with those of \cite{Shannon2018} and \cite{Niu2021}.
%In addition, we find that the non-repeating FRBs are more energetic than the repeating ones as a whole.
%=====================================================================
\begin{deluxetable}{lllc}
%\tablenum{}
\tablecaption{The Best-fit Parameters of the Correlations between $S_{p,obs}$ and fluence for all kinds of FRBs \label{tab2}.}
\tablewidth{0pt}
\tabletypesize{\footnotesize}
\tablehead{
\colhead{FRB}&\colhead{$\gamma$} & \colhead{$\mu$} & \colhead{$\rho^a$}
}
\startdata
Non-repeating \emph{s}FRBs&$0.86\pm0.04$&$0.52\pm0.05$&$0.92$\\
Repeating \emph{s}FRBs    &$0.95\pm0.06$&$0.60\pm0.04$&$0.87$\\
\hline
All \emph{s}FRBs              &$0.86\pm0.03$&$0.55\pm0.03$&$0.92$\\
All \emph{l}FRBs              &$0.71\pm0.12$&$3.50\pm0.06$&$0.88$\\
\enddata
\tablecomments{ \\
$^a$ Pearson correlation coefficient.
}
\end{deluxetable}

In Figure \ref{fig4}, it also can been found that the energies of all FRBs span a broader range from $\sim10^{38}$ to $10^{44}$ erg than those given by \cite{Luo2020} and \cite{Niu2021}. The main reason is that the \emph{l}FRBs with higher fluence (or isotropic energy) but comparable redshift have been included in this study. Besides, we added four new FRBs, including FRBs 181123, 181130, 181118 and 181017.J0036+11 with much lower fluences and larger $DM_{EX}$ measurements detected by FAST.
It is excitingly found that the repeating \emph{s}FRBs/\emph{l}FRBs in our samples are less energetic than non-repeating \emph{s}FRBs/\emph{l}FRBs on a whole. Especially, FRB 121102 has isotropic energy less than $10^{40}$ erg that is good in agreement with \cite{Li2021}. More excitingly, we find from Figure \ref{fig4} that the isotropic energies of \emph{l}FRBs are at least two orders of magnitude larger than those of \emph{s}FRBs. This is quite similar to the energetic difference between short and long GRBs \citep[e.g.][]{Zhangzb2018a}.

\subsection{Peak flux density versus redshift} \label{sec:agn}

\begin{figure*}
\begin{center}
\includegraphics[width=0.6\linewidth]{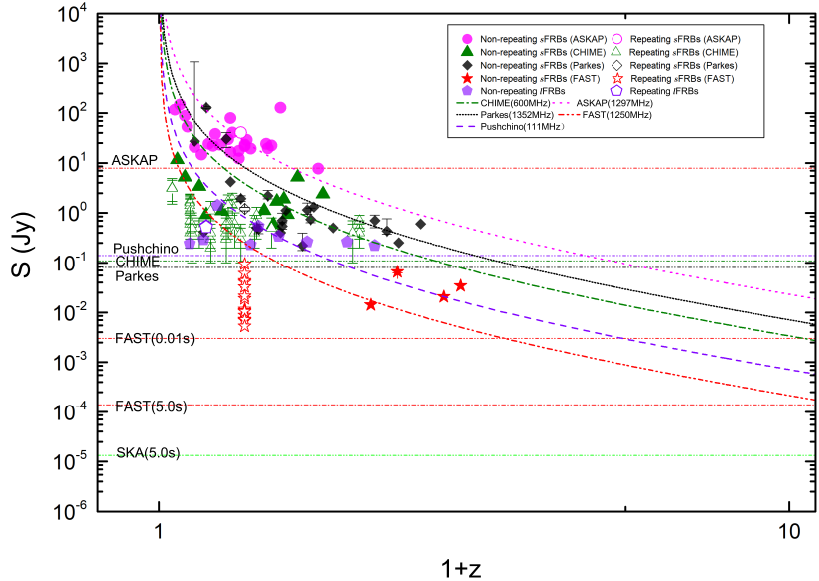}
\end{center}
\caption{Peak flux density vs. redshift for FRBs detected by different telescopes. The slanted curves are plotted for the flux densities evolving with the redshift with an additional negative K-correction effect. The sensitivity limits of the different telescopes are marked by different horizontal lines.
In particular, two sensitivity lines of FAST are given for $\bigtriangleup \tau$=10 ms and $\bigtriangleup \tau$= 5.0 s. The green dotted-dotted-dashed horizontal line represents the sensitivity of SKA at $\bigtriangleup\nu$=700 MHz and $\bigtriangleup\tau$=5 s.
\label{fig5}}
\end{figure*}

In Figure \ref{fig5}, we study the redshift dependence of the
peak flux densities of FRBs based on our samples. We display the evolution profiles of the peak
flux density versus redshift for different telescopes. The isotropic peak luminosity can be determined from $L_p=4\pi D{^2_l}(z)S_{p,obs}(z)\nu_c k$ with a K-correction factor of $k=(1+z)^{\alpha-\beta}$, where $\alpha \sim 0$ and $\beta \sim 1/3$ are the normal temporal and spectral indices defined in $S_{\nu}(t)\propto t^\alpha \nu^\beta$ \citep{Soderberg2004,Chandra2012,Zhangzb2018b}.

%=====================================================================

The sensitivity of different telescopes can be estimated by the equation \citep{Zhangzb2015,Caleb2016}
\begin{equation} \label{e4}
S_{\nu,limt}=\frac{\beta T_{sys} S/N}{ G \sqrt[]{\Delta \nu \Delta \tau N_p}},
\end{equation}
where $T_{sys}$ is the system temperature in unit of K, $S/N$ is the ratio of signal-to-noise, $\beta$ is the digitisation factor, G is the system gain in $K Jy^{-1}$, $\Delta \nu$ is the bandwidth in unit of Hz, $ \Delta \tau$ is the integration time in unit of second, $N_p$ is the number of polarizations. Note that in FRB catalogue the S/N values of \emph{l}FRBs ranging from 7.3 to 10.1 are considerably smaller than those of the majority of \emph{s}FRBs. The S/N differences in burst-detection significance may influence the transparency of our results in a sense. For example, if we apply a S/N threshold of 10 to our analysis, most \emph{l}FRBs would be likely excluded from our sample due to their relatively lower significance level of S/N. According to \cite{Deng2019} and \cite{Josephy2019}, the fluence threshold limits are respectively 2 Jy ms, 51 Jy ms and 7 Jy ms for Parkes, ASKAP and CHIME. In our samples, the maximum pulse widths are 25 ms for Parkes FRB 010312, 6.5 ms for ASKAP FRB 190711 and 63 ms for CHIME FRB 180814.J0422+73. Therefore,one can determine the detection thresholds $S_{\nu,limt}$ of 0.08 Jy, 7.85 Jy and 0.11 Jy for Parkes, ASKAP and CHIME according to Equation \eqref{e4} by adopting the maximum pulse widths instead of the integration times $ \Delta \tau$ of telescopes.

In particular, the sensitivities of FAST are adopted when two following cases are taken into account here. One corresponds to a sensitivity for a maximum pulse width of $\sim10$ ms for \emph{s}FRB 180301 and another one has a sensitivity for a maximum pulse width of $\sim$5.0 s for \emph{l}FRB 160920.
Using the Equation (9) in \cite{Zhangzb2015}, we obtain the detection thresholds $S_{\nu,limt}$ of FAST to be 3.12 $m$Jy for \emph{s}FRBs and 0.14 $m$Jy for \emph{l}FRBs when an aperture efficiency of 0.63 \citep{Jiang2020}, an average system temperature of 24 K \citep{Li2018,Jiang2020}, a signal-to-noise ratio of 7 \citep{Niu2021} and the entire 500-MHz bandwidth are employed. For the ultra-long FRBs detected by Pushchino Radio Astronomy Observatory, the fluctuational sensitivity in a 2.5-MHz receiver bandwidth with a time resolution of 0.1 s is taken as 140 mJy \citep{Fedorova2019a,Fedorova2019b}.

Figure \ref{fig5} displays how peak flux density evolves with redshift. It can be seen that peak flux density will be almost independent of redshift when the distance of a given FRB is far enough. In other words, the detected event rate of FRBs will be approximately independent of redshift, which is very similar to the dependency of peak flux on redshift for the GRBs found by \cite{Zhangzb2018b}.
For comparison, we also estimate the detecting sensitivity of Square Kilometer Array (SKA) by assuming $ \Delta \tau$ = 3.5 s, $\Delta \nu$ = 700 MHz according to \cite{Zhangzb2018b}. Figure \ref{fig5} shows that FAST and SKA exhibit an outstanding detectability for the high-redshift FRBs, which helps us disclose the physical origins of diverse FRBs. It is necessary to be aware of that flux density of FRB host galaxies is surprisingly  independent of redshift as well \citep{Heintz2020}. The consistency of flux density evolving with redshift of FRBs and their hosts are quite similar to that of GRB afterglows and their host galaxies in radio bands \citep{Zhangzb2018b}.
%%%%%%%%%%%%%%%%%%%%%%%%%%%%%%%%%%%%%%%%%%%%%%%%%%%%%%%%%%%%%%%%%%%%%%%%%%%%%%%%
%%%%%%%%%%%%%%%%%%%%%%%%%%%%%%%%%%%%%%%%%%%%%%%%%%%%%%%%%%%%%%%%%%%%%%%%%%%%%%%%
%%%%%%%%%%%%%%%%%%%%%%%%%%%%%%%%%%%%%%%%%%%%%%%%%%%%%%%%%%%%%%%%%%%%%%%%%%%%%%%%
%%%%%%%%%%%%%%%%%%%%%%%%%%%%%%%%%%%%%%%%%%%%%%%%%%%%%%%%%%%%%%%%%%%%%%%%%%%%%%%%

%%%%%%%%%%%%%%%%%%%%%%%%%%%%%%%%%%%%%%%%%%%%%%%%%%%%%%%%%%%%%%%%%%%%%%%%%%%%%%%%
\section{Conclusions and Discussions} \label{sec:conclusions}
Based on the above investigations, our results are summarized as follows:

1. We define our FRB samples with a central frequency below/above 1GHz as subsample I/II, to check the frequency dependence of FRB classifications. It is found that there is a clear
bimodal distribution of the $W_{obs}$ for the subsample I. We separated the subsample I into \emph{s}FRBs and \emph{l}FRBs with a boundary of 100 ms. We find the \emph{s}FRBs with higher central frequency are shorter than those with lower central frequency not only for non-repeating but also repeating \emph{s}FRBs. However, the result of \emph{l}FRBs is ambiguous due to a small number of the samples.
%It is also supported that the non-repeating and repeating short-FRBs have different frequency dependence in view of pulse width and could be thus originated from diverse physical processes.

2. Non-repeating \emph{s}FRBs are brighter than repeating bursts on average. A power law relation of $F \propto S{^\gamma_{p,obs}}$ is found to exist for non-repeating and repeating FRBs with a roughly consistent power-law index of $\gamma\sim0.9$. However, the intercept of the $F-S_{p,obs}$ relations of \emph{l}FRBs and \emph{s}FRBs are largely different.

3. It is confirmed that repeating FRBs in our sample are less energetic than non-repeating \emph{s}FRBs on the whole. More importantly, we find that the averaged isotropic energy of \emph{l}FRBs is larger than that of \emph{s}FRBs at least two orders of magnitude, which is surprisingly similar to the difference between short and long GRBs.

4. In terms of Figure 2 and 3, we conclude that the ``one-off'' events (at least part of them) at present could be repeative in nature, which demonstrates these kinds of repeating and non-repeating FRBs might share the same radiation mechanism. However, \emph{l}FRBs and \emph{s}FRBs are significantly diverse and could originate from different progenitors.

5. We have investigated the dependence of the peak flux density on the cosmological redshift and find that the peak flux density exhibits an independence of the redshift when thedistance of the FRBs becomes farther enough. The ongoing and upcoming large radio telescopes including FAST and SKA have significant potentials to detect more and fainter FRBs at very high redshift in the near future.

In practice, there are various potential factors that misclassify a repeating FRB into an one-off source. The instrumental and analytical biases, such as the beam response and the limited time resolution, lead to some repeating bursts unresolved and missing \citep[e.g.][]{Connor2019,Pleunis2021b}. In addition, the limited sensitivity of current telescopes and the absence of the follow-up observation can also cause the missing of many repeating bursts \citep{Kumar2019,Xiao2021}. These questions are expected to be solved sooner or later, with high-sensitivity multiple facilities and the cumulation of observation time in the era of large telescopes.

Recently, a bright FRB 200428 was reported from a Galactic magnetar \citep{CHIME/FRB2020,Bochenek2020}, of which the discovery confirmed that magnetars are capable of producing FRBs.
\cite{Iwazaki2021} proposed a generation mechanism from the axion star collision with compact
objects to explain some differences between repeating and non-repeating FRBs. In particular, \cite{Iwazaki2021} proposes that the association between the FRB 200428 and magnetar SGR J1935+2154 may result from the collision of an axion with the magnetor. Unfortunately, the number of FRBs produced by an established progenitor including magnetars is too limited nowadays.
Therefore, our statistical results of the observed spectral-temporal properties can impose strong constraints on the generation mechanism of different kinds of FRBs.
%Very recently, Iwazaki (2021) proposed that the axion star collision with neutron star or magnetized accretion disk of galactic black hole generate both repeating and non-repeating FRBs. The generation mechanism can well explain the observed spectral-temporal differences in the repeating and non-repeating FRBs. This difference is a consequence of several factors, including the burst fluence and width distributions, telescope sensitivities, and instrumental selection effects.

The 11 \emph{l}FRBs are all detected at the frequency range of 111 $ \pm$ 2.5 MHz over 6 frequency
channels, each with a narrow receiver bandwidth of 415 kHz \citep{Fedorova2019a, Fedorova2019b}. The unambiguous astrophysical origin is questioned by \cite{Pleunis2021a} because of the low frequency, the narrow receiver bandwidth, the large number of trials in their blind search and the
low S/N value of the claimed events. Meanwhile,
\cite{Nikitin2021} suggested that the effect of the interstellar scatter broadening at 111 MHz low-frequency is stronger than that at higher frequencies and cause the durations of pulse to be several seconds. \cite{Fedorova2019b} suggested that different widths of the \emph{l}FRBs cannot be explained purely by broadening
in the receiver bandwidth and are also associated with internal properties of the pulses.
Interestingly, \cite{Alexander2020} reported a FRB with a wider pulse width of 2.2 s and a specific fluence of 308 Jy ms at an
observing frequency of 111 MHz from SGR J1935+2154 detected by Big Scanning Array of the Lebedev Physical Institute (BSA/LPI), which is interpreted as an off-beam ``slow'' radio burst of the on-beam FRB 200428 associated with SGR burst by \cite{Zhang2021}.
If such an interpretation is correct, the \emph{l}FRBs, at least some of them, could be explained by the off-beam mechanism in the similar way. Consequently, joint observations of full electromagnetic waves (especially at low frequency) and multiple messages including gravitational waves would show new lights on the nature of different kinds of FRBs.
%%%%%%%%%%%%%%%%%%%%%%%%%%%%%%%%%%%%%%%%%%%%%%%%%%%%%%%%%%%%%%%%%%%%%%%%%%%%%%%%
%%%%%%%%%%%%%%%%%%%%%%%%%%%%%%%%%%%%%%%%%%%%%%%%%%%%%%%%%%%%%%%%%%%%%%%%%%%%%%%%
\section{Acknowledgments}
\acknowledgments
We acknowledge the referee for very helpful comments and suggestion. We also thank Prof. Rodin for beneficial discussions on the PUSHCHINO observations of \emph{l}FRBs. This work was supported by National Natural Science Foundation of China (Nos. 11988101, U2031118), the Youth Innovations and Talents Project of Shandong Provincial Colleges and Universities (grant No. 201909118), and the Natural Science Foundations (ZR2018MA030, XKJJC201901, 20165660 and 11104161).

%%%%%%%%%%%%%%%%%%%%%%%%%%%%%%%%%%%%%%%%%%%%%%%%%%%%%%%%%%%%%%%%%%%%%%%%%%%%%%%%
%%%%%%%%%%%%%%%%%%%%%%%%%%%%%%%%%%%%%%%%%%%%%%%%%%%%%%%%%%%%%%%%%%%%%%%%%%%%%%%%

%%%%%%%%%%%%%%%%%%%%%%%%%%%%%%%%%%%%%%%%%%%%%%%%%%%%%%%%%%%%%%%%%%%%%%%%%%%%%%%%
%%%%%%%%%%%%%%%%%%%%%%%%%%%%%%%%%%%%%%%%%%%%%%%%%%%%%%%%%%%%%%%%%%%%%%%%%%%%%%%%

%%%%%%%%%%%%%%%%%%%%%%%%%%%%%%%%%%%%%%%%%%%%%%%%%%%%%%%%%%%%%%%%%%%%%%%%%%%%%%%%
\listofchanges
\end{document}